\documentclass[12pt]{iopart}

\usepackage{graphicx}
\usepackage{hyperref}

\begin{document}

\title[Short-Distance Structure of Nuclei]{Short-Distance Structure of Nuclei}

\author{D W Higinbotham$^1$, E Piasetzky$^2$ and S A Wood$^1$}
\address{$^1$ Thomas Jefferson National Accelerator Facility, Newport News, VA 23606, USA}
\address{$^2$ Tel Aviv University, Tel Aviv 69978, Israel}

\ead{doug@jlab.org}

\begin{abstract}

One of Jefferson Lab's original missions was to further our understanding of the short-distance 
structure of nuclei.  In particular, to understand what happens when two or more nucleons 
within a nucleus have strongly overlapping wave-functions; a phenomena  commonly referred to as short-range
correlations.  Herein, we review the results of the $(e,e^{\prime})$, 
$(e,e^{\prime}p)$ and $(e,e^{\prime}pN)$ reactions that have been used at Jefferson Lab
to probe this short-distance structure as well as provide an outlook for future 
experiments.~\footnote{Document contains copyrighted material that readers may view, browse, and/or download material 
for temporary copying purposes only, provided these uses are for noncommercial personal purposes. 
Except as provided by law, this material may not be further reproduced, distributed, transmitted, 
modified, adapted, performed, displayed, published, or sold in whole or part, without prior 
written permission from the publisher.}

\end{abstract}

\pacs{21.60.-n,24.10.-i,25.30.-c}



%
%

\section{Introduction}

The structure of nuclei is determined by the strong force with 
repulsion at short distances and attraction at moderate distances. 
This force, which binds the nucleons together while also keeping the structure 
from collapsing, makes the nucleus a fairly dilute system. 
This allows for calculations that treat the nucleus as a collection of 
hard objects in an average or mean field to describe many of the properties of nuclear matter;
however, this simple picture breaks down when detailed features are studied.
In particular, the short-distance part of the nucleon-nucleon 
potential can cause central, tensor and spin-orbit correlations
between the nucleons in nuclei which deplete the low momentum states in
nuclear systems~\cite{Rios:2009gb}.  

Calculations indicate that these short-range correlations can lead to short-term 
local densities in the nucleus that are several times as high as the 
average nuclear density of 0.17~GeV/fm$^{3}$; and thus, comparable to densities
predicted in neutron stars~\cite{Frankfurt:2008zv}.  
Isolating the signal 
of short-range correlated nucleons may therefore lead to a deeper 
understanding of cold, dense nuclear systems.

Directly identifying short-range correlations has been experimentally challenging.
The low duty factors and the sub-GeV energies
of previous facilities limited studies of short-range correlations to moderate
momentum-transfer kinematics.   The
Jefferson Lab accelerator and experimental facilities permit measurements
at high energy and high luminosity with a continuous beam.
These parameters have
allowed measurements of the $A(e,e^{\prime})$, $A(e,e^{\prime}p)$ and $A(e,e^{\prime}pN)$
reactions in kinematics above the quasi-elastic peak and at high momentum-transfer.  
These kinematics are where the effects of competing mechanisms such as
meson-exchange currents, isobar configurations, and
final-state interactions are reduced and the signatures of short-range 
correlations become dominate.

%
%

\section{Inclusive Reactions}

\label{src-inclusive}

\begin{figure}[htb]
\includegraphics[width=\textwidth]{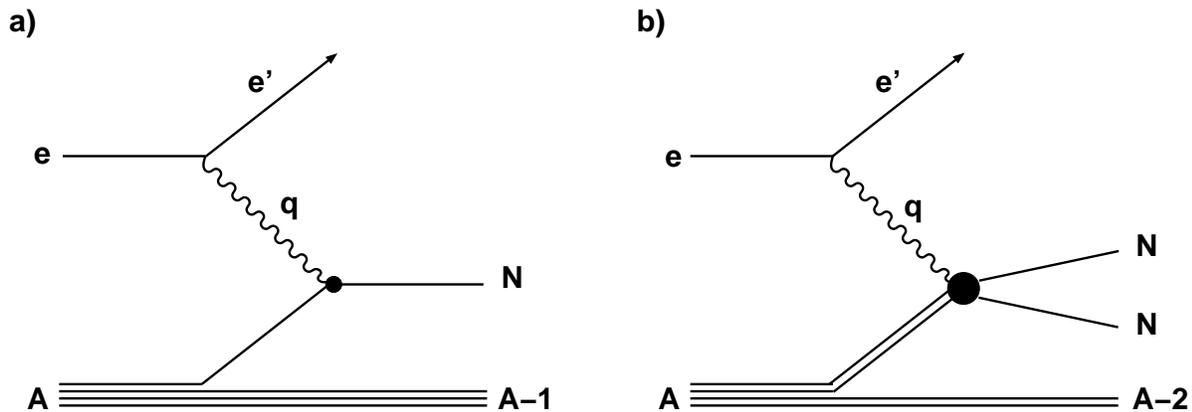}
\caption{\label{src:diagrams}Shown are the dominate diagrams of the $A(e,e^{\prime})$ reaction for x$>$1 and Q$^2 > 1$[GeV/c]$^2$.
Figure a) shows single nucleon scattering while figure b) shows scattering from a correlated initial-state pair.}
\end{figure}

The inclusive $A(e,e^{\prime})$ process is the simplest 
electromagnetic reaction for studying short-range correlations;
though to use this reaction channel, it is important to be in kinematics where
short-range correlations dominate over other reaction mechanisms such 
as meson exchange currents or delta-isobar contributions. 
As first shown with the SLAC $(e,e^{\prime})$ data~\cite{Frankfurt:1993sp}, this is done by requiring
$x_B = Q^{2}/2m\omega > 1$ and $Q^2 > 1~$[GeV/c]$^2$ where $m$ is the nucleon mass, $\omega$ is the energy
transfer and $Q^2$ is the four-momentum transfer squared.  The high $x_B$ minimizes 
the inelastic scattering contributions, while high $Q^2$ minimizes meson-exchange currents.  
It was predicted that in these kinematics diagram b) shown in~\fref{src:diagrams}
would dominate the reaction~\cite{Frankfurt:1981mk}. 
For large $Q^2$ and $x_B>1$ the value of $x_B$ kinematically determines a minimum possible missing momentum;
thus, by choosing $x_B$ large enough, only contributions from nucleons 
with missing momentum above the Fermi sea, corresponding to large excitation 
energies, can contribute to the reaction.

For a given $Q^2$ range, the SLAC inclusive data was plotted as function of $x_B$ for the weighted
ratio of yields, 
\begin{eqnarray}
r(A,{}^2\hbox{H}) = \frac{{\cal Y}(A)}{Z\sigma_{ep} + N\sigma_{en}} \frac{\sigma_{ep}+ \sigma_{en}}{{\cal Y}(^2\hbox{H})} 
C^A_{\rm rad},
\label{src-equation}
\end{eqnarray}
where  $\cal Y$ is the normalized yield, $Z$ and $N$ are the number of protons and neutrons in a nucleus
$A$, $\sigma_{eN}$ is the electron-nucleon cross section,
and $C_{\rm rad}^A$ is the ratio of the radiative correction factors
for $A$ and $^2$H.
The SLAC data show for $Q^2 > 1 [GeV/c]^2$ and $x_B > 1.4$ a scaling in the data which 
was argued to be due to short-range correlations~\cite{Frankfurt:1993sp}.  These
events corresponded to nucleon momenta of about 250 - 300 MeV/c.
For these measurements the statistics were rather limited and $x_B$ was less than 2. 
Also, the deuteron data was measured in a  different kinematics and comparisons to the other 
nuclei required nontrivial extrapolations.

\begin{figure}[htb]
\includegraphics[width=\textwidth]{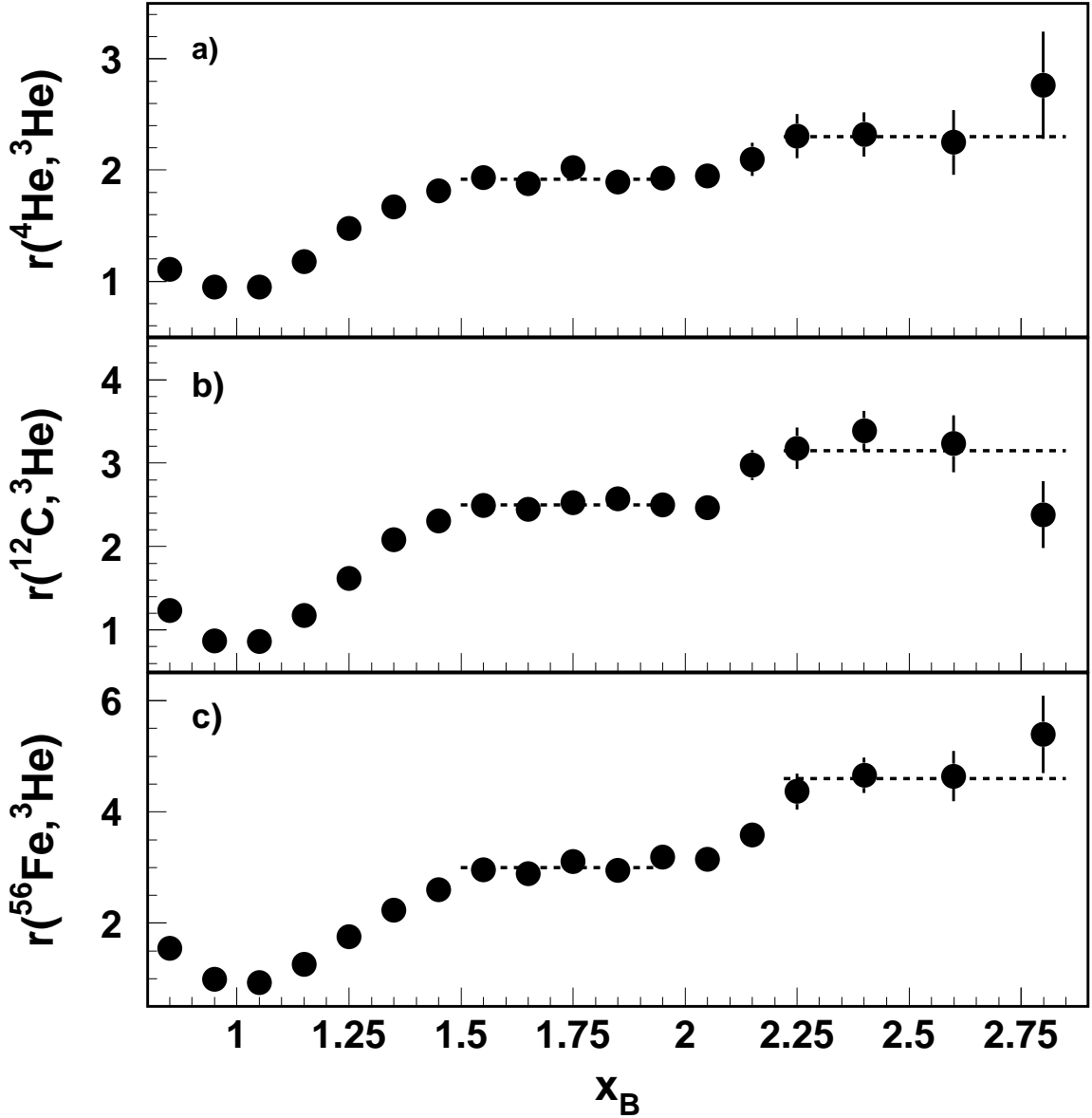}
\caption{\label{src:inclusive-ratios}Weighted cross section ratios of 
(a) $^4$He, (b) $^{12}$C and (c) $^{56}$Fe to $^3$He as a function of
$x_B$ for $Q^2>1.4$ GeV$^2$.  The horizontal  dashed lines indicate the two-nucleon 
and three-nucleon scaling regions used to calculate the
per-nucleon probabilities for two- and three-nucleon short-range correlations
in nucleus $A$ relative to $^3$He.  
Reprinted with permission from
\href{http://dx.doi.org/10.1103/PhysRevLett.96.082501}{Egiyan K~S {\em et~al.\/} (CLAS) 2006 {\em Phys. Rev. Lett.\/} {\bf 96} 082501.}
Copyright 2006 by the American Physical Society.
}
\end{figure}

In order to improve upon the SLAC $(e,e^{\prime})$ data,
Jefferson Lab provided higher statistics data and was able to 
look for a second scaling region due to three-nucleon correlations.  The three-nucleon correlation
region was accessed  by taking the ratios to $^{3}$He data instead of deuterium. 
The data was taken with the CEBAF Large Acceptance Spectrometer (CLAS)~\cite{Egiyan:2003vg,Egiyan:2005hs}
and is shown in~\fref{src:inclusive-ratios}.
The new data not only show scaling in the $1 > x_B > 2$ region; but also show, for
the first time, a second scaling in the $x_B > 2$ region.
This second scaling has been interpreted as being due to
three-nucleon correlations and further strengthens the argument that the first scaling region is due to
short-range two-nucleon correlations~\cite{Strikman:2005wu}.

From these ratios, one can determine the probabilities of two-nucleon correlations 
in the various nuclei by using the relatively well understood deuteron.
This is valid due to closure in the interaction of the knocked-out nucleons with nucleons not belonging
to correlations and cancellation of final-state interactions within the correlations in the ratio~\cite{Frankfurt:1981mk}.
The probability of a nucleon-nucleon correlation in deuterium is defined as the probability that a
nucleon in deuterium has a momentum $p > 275$~MeV/c.  These are  
momenta that have been shown in $\vec{D}(\vec{e},e^{\prime}p)$ asymmetry
measurements to be dominated highly correlated nucleons~\cite{passchier:2001uc}. 
Thus by integrating the wave function, it has been found that 
the per-nucleon probability of a  nucleon in deuterium, $a_2 (^2H)$, to be in a nucleon-nucleon
correlation is is $0.041\pm0.008$.    
The probability of correlations in deuterium versus $^3$He, $a_2({}^3\hbox{He}/{}^2\hbox{H})$ has been determined to
be $1.97\pm0.1$ and thus $a_{2N}({}^3\hbox{He}) = 0.08\pm 0.016$.  From this result, the general $a_{2N}(A)$
probabilities can be calculated from the scaling results for $1.15 < x_B < 2$ for $r(A/{}^3\hbox{He}) \times
r({}^3\hbox{He}/{}^2\hbox{H})$.
To obtain the absolute probability of three-nucleon short-range correlations,
a calculation of the Bochum group~\cite{Nogga:2002qp} using various potentials
was used and gave an average value of 
$a_{3N}({}^3\hbox{He})=0.18\pm0.06\%$.  From this it was concluded
that the $3N$ short-range correlations must be less then 1\%~\cite{Egiyan:2005hs}.
These results are summarized in~\tref{src:inclusive-table}.

\begin{table}[htb]
\caption{\label{src:inclusive-table}
Shown are the average ratios, $r(A/{}^3\hbox{He})$, for the scaling 
regions $1.5<x_B<2$ and $2.2<x_B<2.7$
along with the extracted absolute probabilities $a_{2N}(A)$ and $a_{3N}(A)$ that in a nucleus $A$
a two- or three-nucleon short-range correlation is taking place at a given instant. 
Statistical and systematic errors have been combined in quadrature.
}
\begin{indented}
\item[]\begin{tabular}{@{}lcccc}
\br
          &$r(A/{}^3\hbox{He})$ & $a_{2N}(A)$  & $r(A/{}^3\hbox{He})$  &$a_{3N}(A)$ \\ 
	  & $1.5<x_B<2$	        & [\%]	              &	 $2.2<x_B<2.7$     &  [\%]              \\ \mr
$^{3}$He  & 1                   &  8.0$\pm$1.6     & 1                     & 0.18$\pm$0.06   \\ 
$^{4}$He  & 1.93$\pm$0.03       & 15.4$\pm$3.2     & 2.33$\pm$0.13         & 0.42$\pm$0.14   \\ 
$^{12}$C  & 2.49$\pm$0.15       & 19.8$\pm$4.4     & 3.18$\pm$0.27         & 0.56$\pm$0.21   \\ 
$^{56}$Fe & 2.98$\pm$0.18       & 23.9$\pm$5.3     & 4.63$\pm$0.33         & 0.83$\pm$0.27   \\ 
\br
\end{tabular}
\end{indented}
\end{table}

In Jefferson Lab's Hall~C, inclusive data was also taken for $x_B > 1$ and $Q^2$ from 1-7 [GeV/c]$^2$.  
This data also showed, for initial nucleon momenta greater then 300~MeV/c, 
a scaling behavior consistent with short-range nucleon-nucleon correlations~\cite{Arrington:1998ps}. 
Recent calculations show that these scaling regions can be used to obtain useful information about short-range 
correlations~\cite{CiofidegliAtti:2009qc}


\section{Proton Knock-out}
\label{src-knock-out}

For proton knock-out experiments, it is useful to introduce the concept of
an initial state momentum distribution of nucleons within the nucleus.   This
is done in a factorized approximation with a spectral function $S({\bf p},E)$ which relates the experimental 
missing momentum, {\bf p}, and missing energy, E to the measured
$(e,e^{\prime}p)$ cross sections by the relation:
\begin{eqnarray}
\frac{d^6\sigma}{dE\,dE^{\prime}\,d\Omega_e\,dE_p\,d\Omega_p}
   = K\,\sigma_{eN} S({\bf p},E)\,T_A(Q^2)
\label{spectralfunction}
\end{eqnarray}
where $K$ is a kinematic factor, $\sigma_{eN}$ is the elementary cross
section, and the transparency, $T_A$, 
acts as the normalization factor between the experimental cross section
and a theoretical momentum distribution.
The transparency is often interpreted as the probability that a 
nucleon will be emitted from the nucleus without rescattering, though other effects
are included in this factor, as it is experimentally just the scale parameter required 
to reconcile a theoretical momentum distribution
with an experimental cross sections.  Having theory calculate both the spectral function
and the transparency, allows us to further test our understanding of nuclear matter in a many-body
system where exact solutions are not yet calculable.


In independent particle models, the strength 
of the spectral function is almost
entirely limited to momenta and energies less than $250~{\rm MeV}/c$ and
$80~{\rm MeV}$, namely the Fermi momentum and energy. In models built
up from realistic nucleon-nucleon potentials, which include a
repulsive core and medium range tensor component, the
spectral function strength in the Fermi-gas region is reduced 
by about 20\%~\cite{Benhar:1989,Benhar:2005dj}.
Measurements with reactions such as $(e,e^{\prime}p)$ show a reduction of at
least 30\%~\cite{Lapikas:1993,Kelly:1996hd}, which is 
consistent with the expectation that short-range
correlations are an important part of the description of the nucleus.
In the Hall~A $^{16}$O$(e,e^{\prime}p)$ experiment,
which measured cross sections and response functions
for $25 < E < 120$~MeV and $p < 340$~MeV/c, it was found that calculations
that included pion exchange currents, isobar currents and two-nucleon correlations
were required to account for the shape of the data at high missing momentum, 
though the calculations still under estimated the measured cross section 
by a factor of two~\cite{Liyanage:2000bf,Fissum:2004we}.

Measurements of $A(e,e^{\prime}p)$ were made in Jefferson Lab's Hall~C 
to directly identify correlated strength rather than inferring it from
the absence of strength predicted by independent particle models.  
Using the HMS and SOS spectrometers, protons knocked out of $^{12}$C
were detected in parallel 
kinematics with missing energy $E>40~{\rm MeV}$ and missing momentum
$k>240~{\rm MeV}/c$.  The spectral function derived from this
measurement, integrated over this range of missing energy and
momentum, accounts for approximately {10\%} of the total strength of
scattering from protons in 
$^{12}$C~\cite{Rohe:2004dz}.
This strength is consistent with or larger than the correlated
strength predicted for this kinematic range from a Correlated Basis
Function approach~\cite{Benhar:1989} or Green's function approach to
calculating the spectral function~\cite{Muther:1995bk}.  
These calculations both predict
more of the correlated strength to be outside of the measured range,
including at low missing energy and momentum where strength is
indistinguishable from IP model proton knockout.

At Jefferson Lab, $T_A$ has been measured for
several nuclei~\cite{Abbott:1997bc,Garrow:2001di} and found
to be essentially independent of proton energy for knocked-out protons
with over 1 GeV of kinetic energy.  The measured transparencies have
been found to be somewhat higher than expected for a proton
propagating through a independent particle model nucleus.  For example, for $^{12}$C
the measured $T_A = 0.59$, while the prediction from an
independent particle shell model is 0.55~\cite{Rohe:2005vc}.  
Calculations with the Correlated Basis Function approach increase the transparency by about
10\% due to short-range correlations and bring experiment and theory into 
much better agreement.  This is interpreted as the short-range  nucleon-nucleon
causing a depletion in density near the point of interaction, giving a lower overall probability
that is will interact as it leaves the nucleus.

\begin{figure}[htb]
\includegraphics[width=\textwidth]{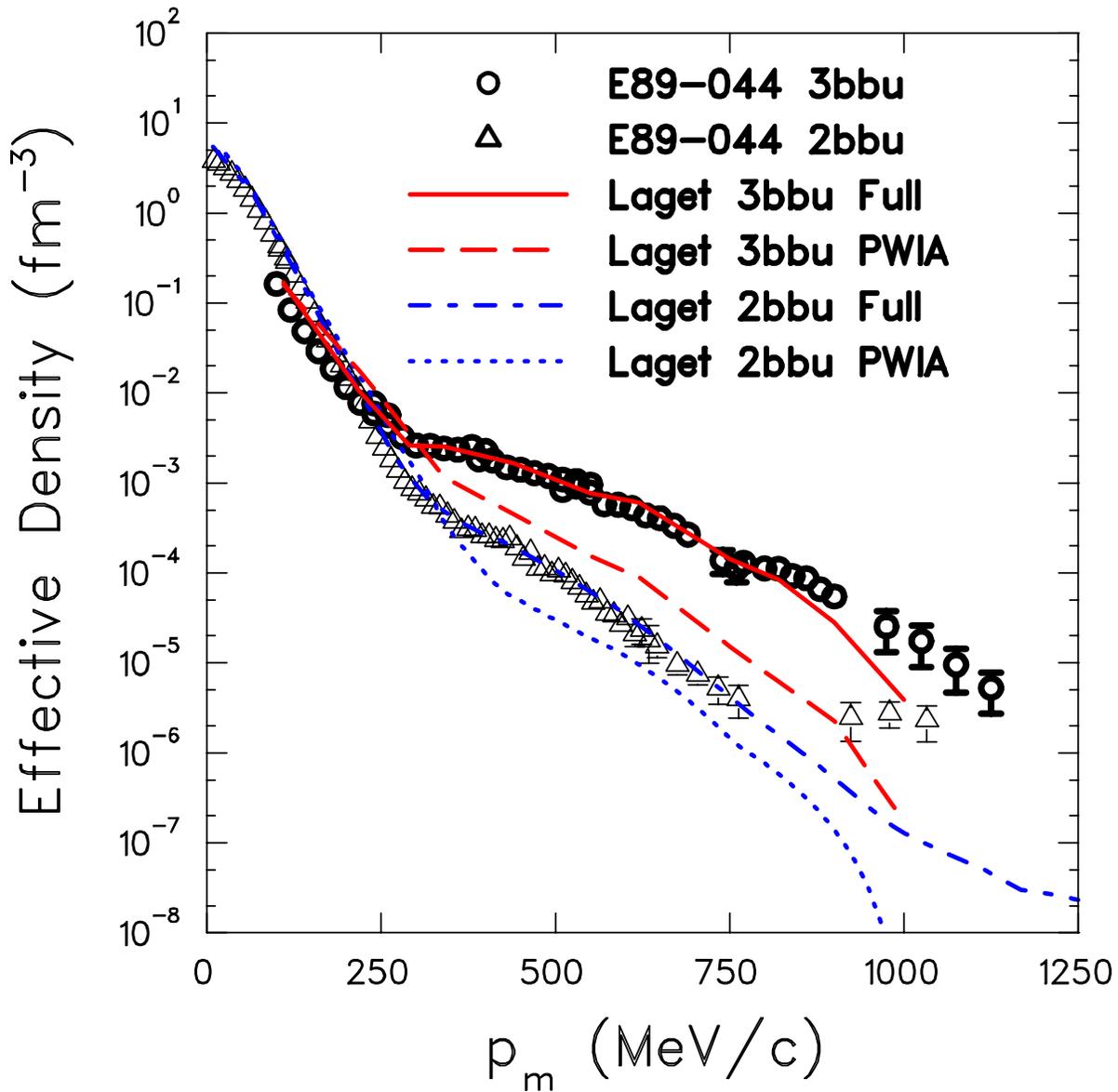}
\caption{\label{src:he3density}Proton effective momentum density distributions in $^3$He 
extracted from $^3$He$(e,e^{\prime}p)pn$ three-body break-up (3bbu) is shown as the
open black circles and the $^3$He$(e,e^{\prime}p)d$ two-body break-up (2bbu) is shown
as open black triangles.  
The three-body break-up (3bbu) integration covers $E_M$ from threshold to 140~MeV. 
The results are compared to calculations from J.-M.~Laget~\cite{Laget:2004sm} which explain the
dominance of the continuum cross section at large missing momentum as a strong interference 
between short-range correlations and final-state interactions.  
Reprinted with permission from 
\href{http://dx.doi.org/10.1103/PhysRevLett.94.082305}{Benmokhtar F {\em et~al.\/} (Hall A) 2005 {\em Phys. Rev. Lett.\/} {\bf 94} 082305.}
Copyright 2005 by the American Physical Society.}
\end{figure}

Hall~A measurements of the $^3$He$(e,e^{\prime}p)d$ and $^3$He$(e,e^{\prime}p)pn$
reactions were taken with a beam energy of 4.8~GeV, $Q^2$ = 1.5~[GeV/c]$^2$ in $x_B = 1$ kinematics.
When this experiment was proposed, it was expected that these kinematics would cleanly show short-range 
correlations at missing momenta greater then 300~Mev/c.  What was observed was a much greater strength
in the high missing momentum region then expected~\cite{Benmokhtar:2004fs,Rvachev:2004yr}
as shown in figure~\ref{src:he3density}.
The three-body break-up (3bbu) has been integrated over missing energy so that it could 
be plotted with the two-body break-up (2bbu) and the strengths of the two reactions compared.

The unexpected strength was explained as an interference between correlations
in the initial state and final-state interaction~\cite{Laget:2004sm}, where neither effect alone could 
explain the observed cross section.  Subsequently calculations by Ciofi degli Atti and Kaptari
 showed that a parameter free calculation which included a realistic wave-function~\cite{Wiringa:1994wb}, 
i.e. one in which correlations are built in,
and final-state interactions calculated with the generalized Eikonal approximation could explain the 
data~\cite{degliAtti:2005dh,CiofidegliAtti:2005qt,CiofidelgiAtti:2007qu}.

%
%

\section{Triple Coincidence Reactions}
\label{src:triple-coincidence-reactions}

While the $A(e,e^{\prime})$ and $A(e,e^{\prime}p)$ data shown 
in sections~\ref{src-inclusive} and~\ref{src-knock-out}
clearly suggest strong local correlations, it is the new exclusive data that  confirms that the 
inclusive scaling is indeed due to short-range correlations.   Also, with exclusive reactions,
it possible to determine types and abundances of nucleon pairs
involved in correlations.  Such experiments where two nucleons are detected
were proposed in the 1960's~\cite{Yu1966392}, but only with the advent of
high luminosity and high energy machines were clean kinematics (i.e. $Q^2 > 1$ and $x_B > 1$) obtainable.

\begin{figure}[hbt]
\includegraphics[width=\textwidth]{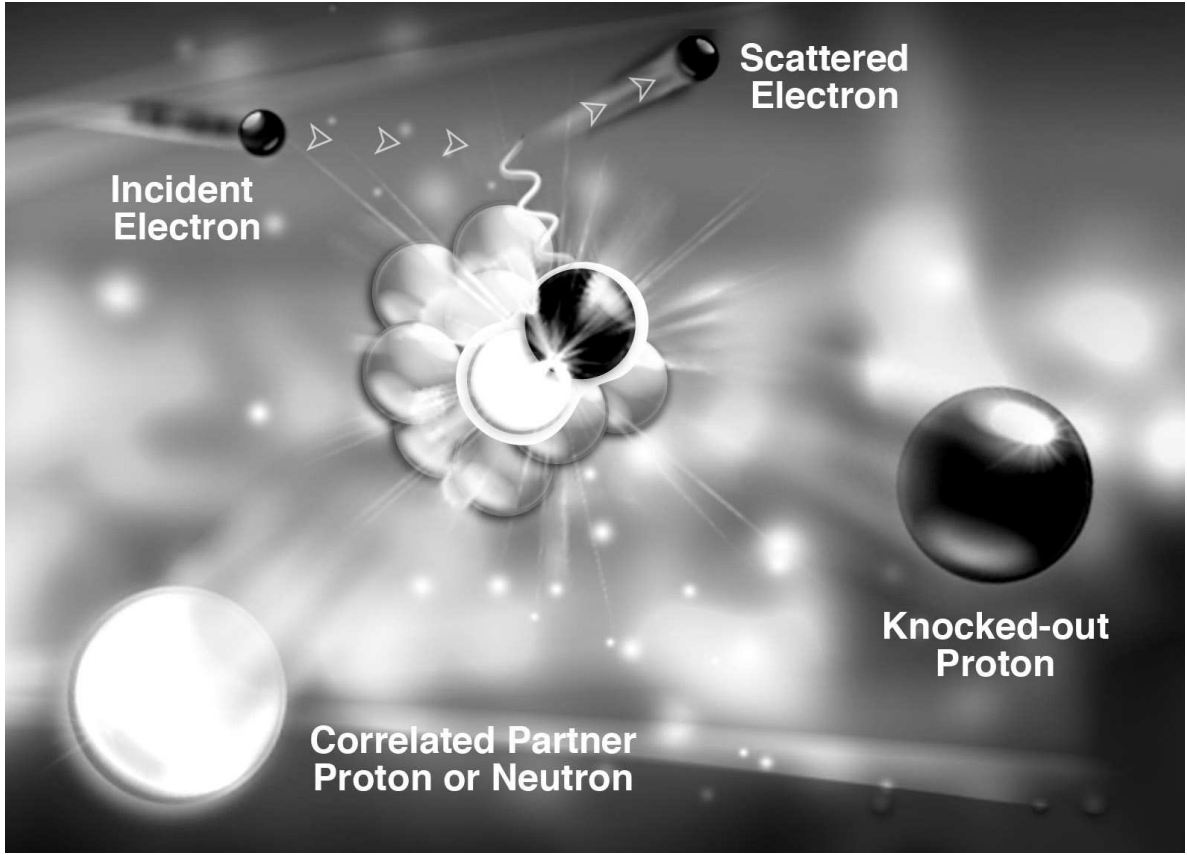}
\caption{\label{src-cartoon-bw}Illustration showing the $(e,e^{\prime}pN)$ reaction.  The incident
electron couples to a nucleon-nucleon pair via a virtual photon.  In
the final state, the scattered electron and struck nucleon
are detected along with the
correlated nucleon that is ejected from the nucleus.  Illustration courtesy of Joanna Griffin and inspired by Anna Shneor.}
\end{figure}

Experimentally, a high-momentum, small de Broglie wavelength probe,
can knock-out one nucleon of a nucleus while leaving the residual nucleus nearly unaffected; 
but if the nucleon being struck is part of an initial-state pair, the high
relative momentum of the pair will cause the correlated nucleon to
recoil and be ejected~\cite{Frankfurt:1981mk,Frankfurt:1988nt}.
Figure~\ref{src-cartoon-bw} shows an illustration of a high-energy electron
knocking-out two nucleons from a nucleus.

Historically, the interpretation of triple-coincidence experimental 
data at low $Q^2$ in terms of short-range correlations has been complicated 
by contributions from meson-exchange currents,
isobar configurations and final-state interactions~\cite{Kester:1995zz,Blomqvist:1998gq,Onderwater:1998zz,Groep:2000cy}.
The kinematics for the measurements described herein were chosen to
minimize these effects.  For example, at high $Q^2$, meson-exchange contributions
decrease as $1/Q^2$ relative to plane-wave impulse contributions and
relative to those due to correlations~\cite{Arnold:1989qr,Laget:1987hb}.  
Large $Q^2$ and $x_B$ also drastically reduce isobar currents 
contributions~\cite{Frankfurt:1996xx,Sargsian:2001ax}. Final-state interactions are not
small; but, in the chosen kinematics, predominately restricted to be
within the correlated pair~\cite{Frankfurt:2008zv}. 

The high luminosity Hall~A triple coincidence experiment used
an incident electron beam of 4.627 GeV and 
the two Hall A high-resolution
spectrometers (HRS)~\cite{Alcorn:2004sb} to identify the
$^{12}$C$(e,e^{\prime}p)$ reaction.  
Scattered electrons were detected
in the left HRS at a central scattering angle (momentum) of
19.5$^{\circ}$ (3.724 GeV/c).
This corresponds to the quasi-free
knockout of a single proton with transferred three-momentum $| \vec{q}
|$ = 1.65 GeV/c, transferred energy $\omega=0.865$ GeV, $Q^2=2$
(GeV/c)$^2$, and $x_{B} = 1.2$.
Knocked-out protons were detected using the right
HRS which was set at 3 different combinations of central angle
and momentum: 40.1$^{\circ}$ \& 1.45 GeV/c, 35.8$^{\circ}$ \& 1.42
GeV/c, and 32.0$^{\circ}$ \& 1.36 GeV/c.  
These kinematic settings
correspond to median missing-momentum values $p_{miss}$ = 0.35, 0.45
and 0.55 GeV/c, respectively; covering the $p_{miss}$ range of 300
-- 600 MeV/c.

A large-acceptance spectrometer, BigBite, was used to detect
recoiling protons in $^{12}$C$(e,e^{\prime}pp)$ events.  The
BigBite spectrometer consists of a large-acceptance,
non-focusing  dipole magnet~\cite{deLange:1998au} and a custom detector package.
For this measurement, the magnet was located at an angle of 99$^{\circ}$ and
1.1~m from the target with a resulting angular acceptance of about
96~msr and a nominal momentum acceptance from 0.25 GeV/c to
0.9~GeV/c.  Located immediately behind BigBite was a large neutron array
matched to BigBite's acceptance.  This made it possible to simultaneously detect
the $^{12}C(e,e^{\prime}pp)$ and $^{12}C(e,e^{\prime}pn)$ reactions.

\begin{figure}
\includegraphics[width=\linewidth]{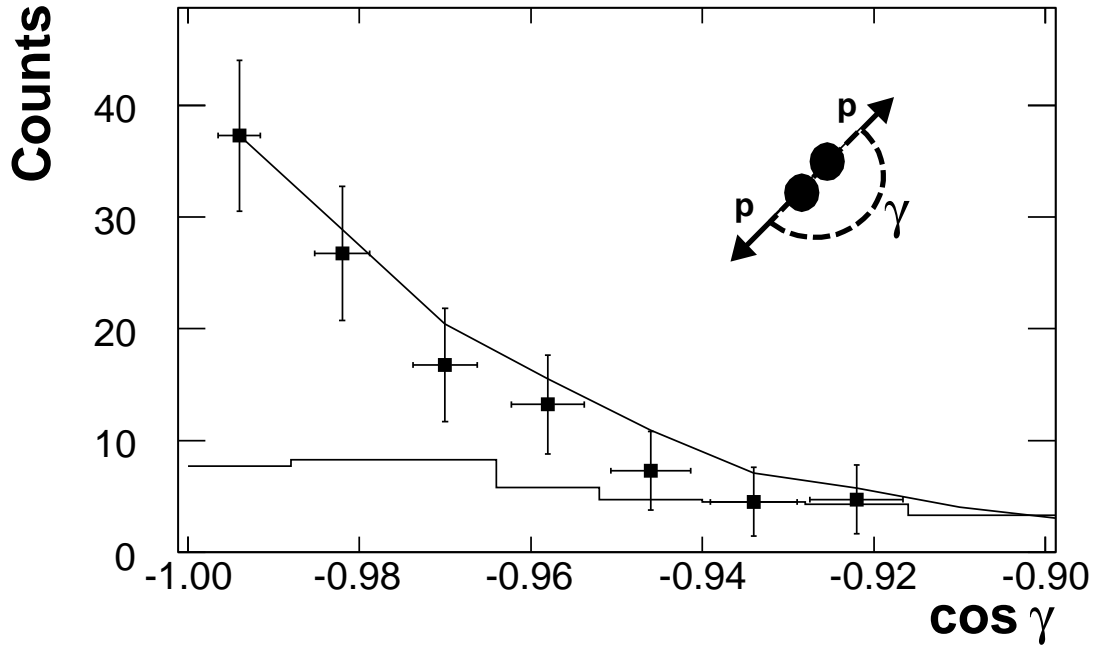}
\caption{\label{src:bigbitecosine}For the 
$^{12}$C$(e,e^{\prime}pp)$
reaction at $Q^2 > $~[1GeV/c]$^2$, 
the distribution of the cosine of the opening angle between the
$\vec p_{miss}$ and $\vec p_{rec}$ for a $p_{miss}=0.55$~GeV/c.
The histogram shows the distribution of random
events.  The curve is a simulation of the scattering off a moving pair
with a width of 0.136~GeV/c for the pair center of mass momentum.  
Reprinted with permission from 
\href{http://dx.doi.org/10.1103/PhysRevLett.99.072501}
{Shneor R {\em et~al.\/} (Hall A) 2007 {\em Phys. Rev. Lett.\/} {\bf 99} 072501.}
Copyright 2007 by the American Physical Society.}
\end{figure}

\Fref{src:bigbitecosine} shows the cosine of the
angle, $\gamma$, between the missing momentum ($\vec p_{miss}$) and
the recoiling proton detected in BigBite ($\vec p_{rec}$) for the
highest $p_{miss}$ setting of 550~MeV/c~\cite{Shneor:2007tu}. 
Also shown in \fref{src:bigbitecosine} is the angular correlation for the random background as
defined by a time window offset from the coincidence peak.  The back-to-back
peak of the real triple coincidence events is demonstrated clearly.
The curve is a result of a simulation of the scattering off a moving
pair having a center-of-mass momentum width of 0.136~GeV/c. That
width was extracted from the data and is consistent with the width
measured in the $(p,ppn)$ experiment at Brookhaven National Lab~\cite{Tang:2002ww}
and well as with a theoretical calculation based on
the convolution of two independent single particle momentum distributions~\cite{CiofidegliAtti:1995qe}.

\begin{figure}[htb]
\includegraphics[width=\linewidth]{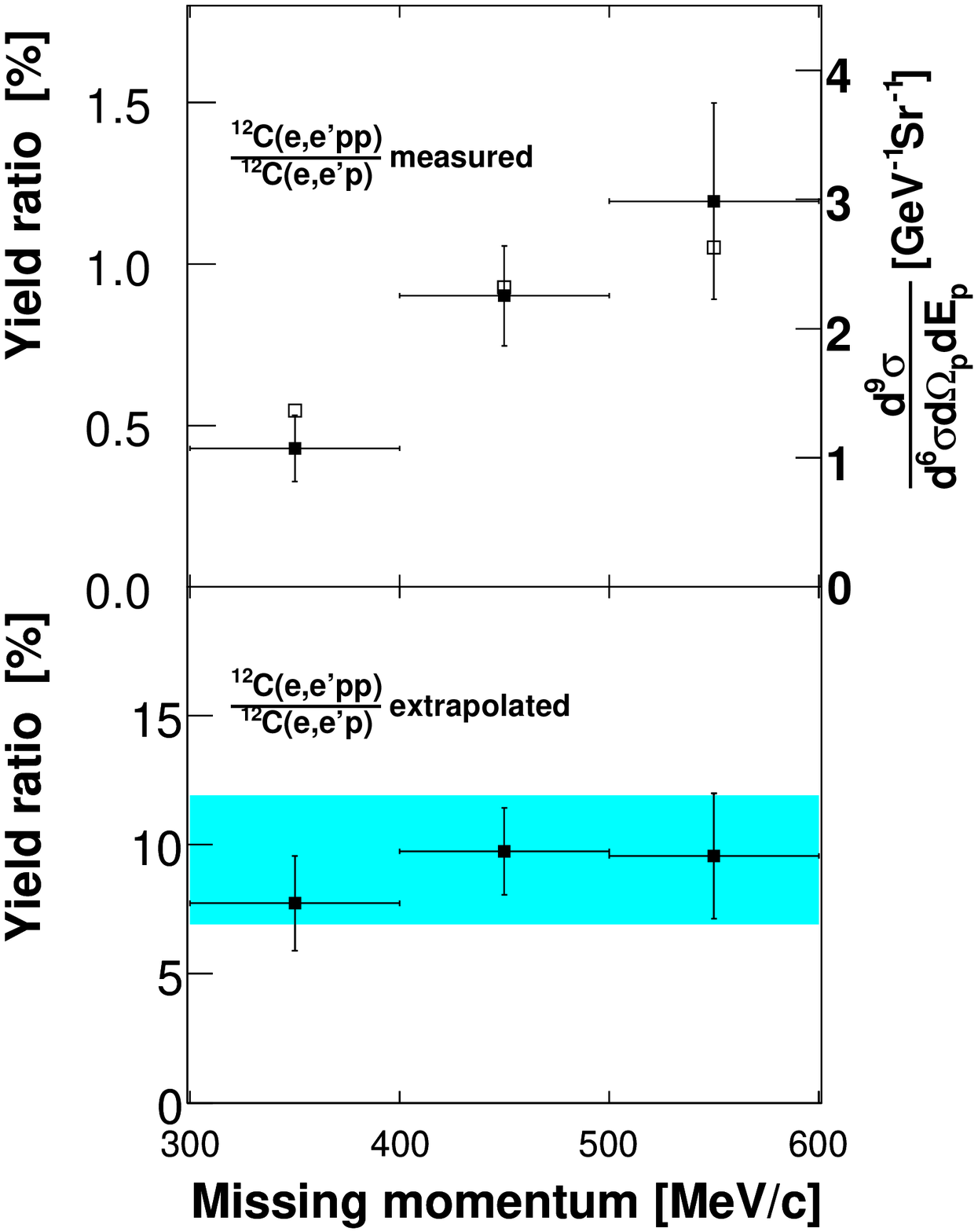}
\caption{\label{src:yieldratios}The measured and extrapolated ratios of yields for the
$^{12}$C(e,e$^{\prime}$pp) and the $^{12}$C(e,e$^{\prime}$p)
reactions.  The shaded area represents a band of $\pm 2 \sigma$
uncertainty in the width of the center-of-mass momentum of the pair. 
Reprinted with permission from 
\href{http://dx.doi.org/10.1103/PhysRevLett.99.072501}
{Shneor R {\em et~al.\/} (Hall A) 2007 {\em Phys. Rev. Lett.\/} {\bf 99} 072501.}
Copyright 2007 by the American Physical Society.}
\end{figure}

The measured ratio of $^{12}$C$(e,e^{\prime}pp)$ to $^{12}$C$(e,e^{\prime}p)$
events is given by the
ratio of events in the background-subtracted  time-of-flight peak
to those in the  $^{12}$C(e,e$^{\prime}$p) spectra. The measured
ratios are limited by the finite acceptance of BigBite.  The
center of mass momentum distribution obtained from the measured angular
correlation is used to account for this finite acceptance;
the resulting extrapolated ratios are shown in~\fref{src:yieldratios}. 

To compare the
$(e,e^{\prime}pn)$ to $(e,e^{\prime}p)$
yields, a similar procedure as
above was followed.
With neutrons, momentum determination was done with time-of-flight and
the neutron detection efficiency was only about 40\%.  This results
in a poorer signal to background ratio than the $(e,e^{\prime}pp)$ measurement and
also a larger uncertainty.  Taking into account the finite acceptance of the
neutron detector and the neutron detection efficiency,
it was   found that $96 \pm 22$\% of the $(e,e^{\prime}p)$ events with a missing momentum
above 300~MeV/c had a
recoiling neutron.   This result agrees with the hadron beam measurement
of the $(p,2pn)/(p,2p)$ reaction
in which $92 \pm 18$\% of the $(p,2p)$ events with a missing momentum
above the Fermi momentum of 275~MeV/c were found to have a single
recoiling neutron carrying the momentum~\cite{Piasetzky:2006ai}.

Since the recoiling proton
$^{12}$C$(e,e^{\prime}pp)$ and neutron $^{12}$C$(e,e^{\prime}pn)$
data were collected simultaneously with detection systems covering
nearly identical solid angles, the ratio of 
$^{12}$C$(e,e^{\prime}pn)$/$^{12}$C$(e,e^{\prime}pp)$
could be directly determined with many of
the systematic factors needed to compare the rates of the
$^{12}$C(e,e$^{\prime}$pp) and $^{12}$C(e,e$^{\prime}$pn)
reactions canceling out.
Correcting only for detector efficiencies,
a ratio of $8.1\pm 2.2$ was determined.  To estimate the effect of final-state
interactions (i.e., reactions that happen after the initial
scattering), 
the attenuation of the recoiling protons
and neutrons was assumed to be almost equal.  In this case, the only correction
related to final-state interactions of the measured 
$^{12}$C$(e,e^{\prime}pn)$ to $^{12}$C$(e,e^{\prime}pp)$
ratio is due to single charge exchange.  
For the experiment, the neutron to proton single charge exchange 
dominates proton to neutron single charge exchange
since the $(e,e^{\prime}pn)$ rate is about an order of magnitude larger than the
$(e,e^{\prime}pp)$ rate and thus decreases the measured
$^{12}$C$(e,e^{\prime}pn)$/$^{12}$C$(e,e^{\prime}pp)$
ratio.
Using Glauber
approximation~\cite{Mardor:1992sb}, it was estimated this effect was
11\%. Taking this into account, the corrected experimental ratio for
$^{12}$C$(e,e^{\prime}pn)$/$^{12}$C$(e,e^{\prime}pp)$
is $9.0\pm 2.5$.

\begin{figure}[htb]
\vspace{0.4cm}
\includegraphics[angle=-90,width=\textwidth]{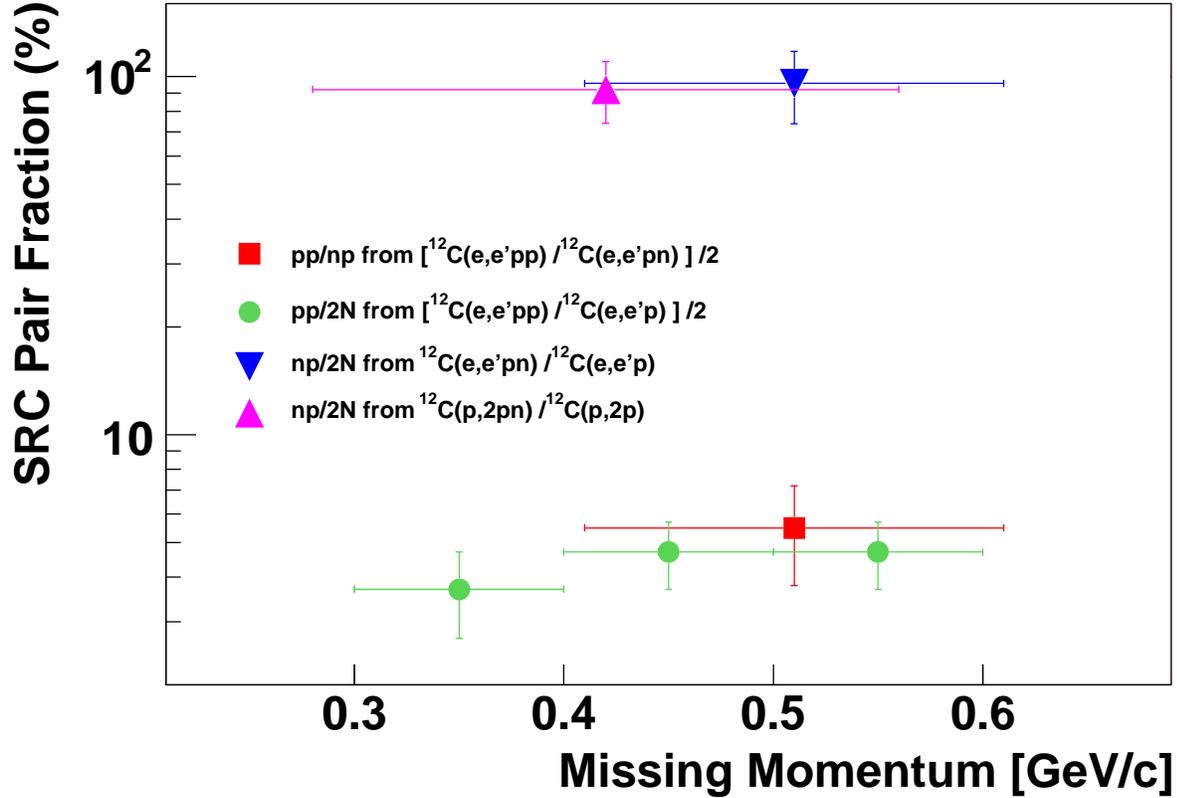}
\caption{\label{src:pairfraction}The fractions of correlated pair combinations in carbon as obtained from
the 
$(e,e^{\prime}pp)$ to $(e,e^{\prime}pn)$
reactions~\cite{Subedi:2008zz},
as well as from previous $(p,2pn)$ data~\cite{Piasetzky:2006ai}. 
From 
\href{http://dx.doi.org/10.1126/science.1156675}{Subedi R {\em et~al.\/} (Hall A) 2008 {\em Science\/} {\bf 320} 1476.}
Reprinted with permission from AAAS.}
\end{figure}

To deduce the ratio of $pn$ to $pp$ correlated pairs in the ground state of
$^{12}$C,  the measured 
$^{12}$C$(e,e^{\prime}pn)$ to $^{12}$C$(e,e^{\prime}pp)$
ratio was used.  Since the experiment triggered only on forward
$(e,e^{\prime}p)$ events,
the probability of detecting $pp$ pairs was twice that of $pn$
pairs; thus, we conclude that the ratio of $pn/pp$ pairs in the
$^{12}$C ground state is $18\pm5$ as shown in~\fref{src:pairfraction}.
This result is consistent with the $(p,2pn)$ data~\cite{Piasetzky:2006ai}.  Since both
the probe and the kinematics of these two experiment were very different, it furthers
the interpretation of the process as being due to scattering off a correlated pair of nucleons~\cite{Higinbotham:2009zz}.

The small ratio of $pp/np$ has been explained by several theoretical
groups~\cite{Schiavilla:2006xx,Sargsian:2005ru,Alvioli:2007zz}.   These
calculations clearly show
that the measured $pp/np$ ratio is expected and is a clear
indication of the nucleon-nucleon tensor force at the probed distances and relative
momenta of the nucleons in a short-range two-nucleon correlation. 

\begin{figure}[t]
\includegraphics[width=\textwidth]{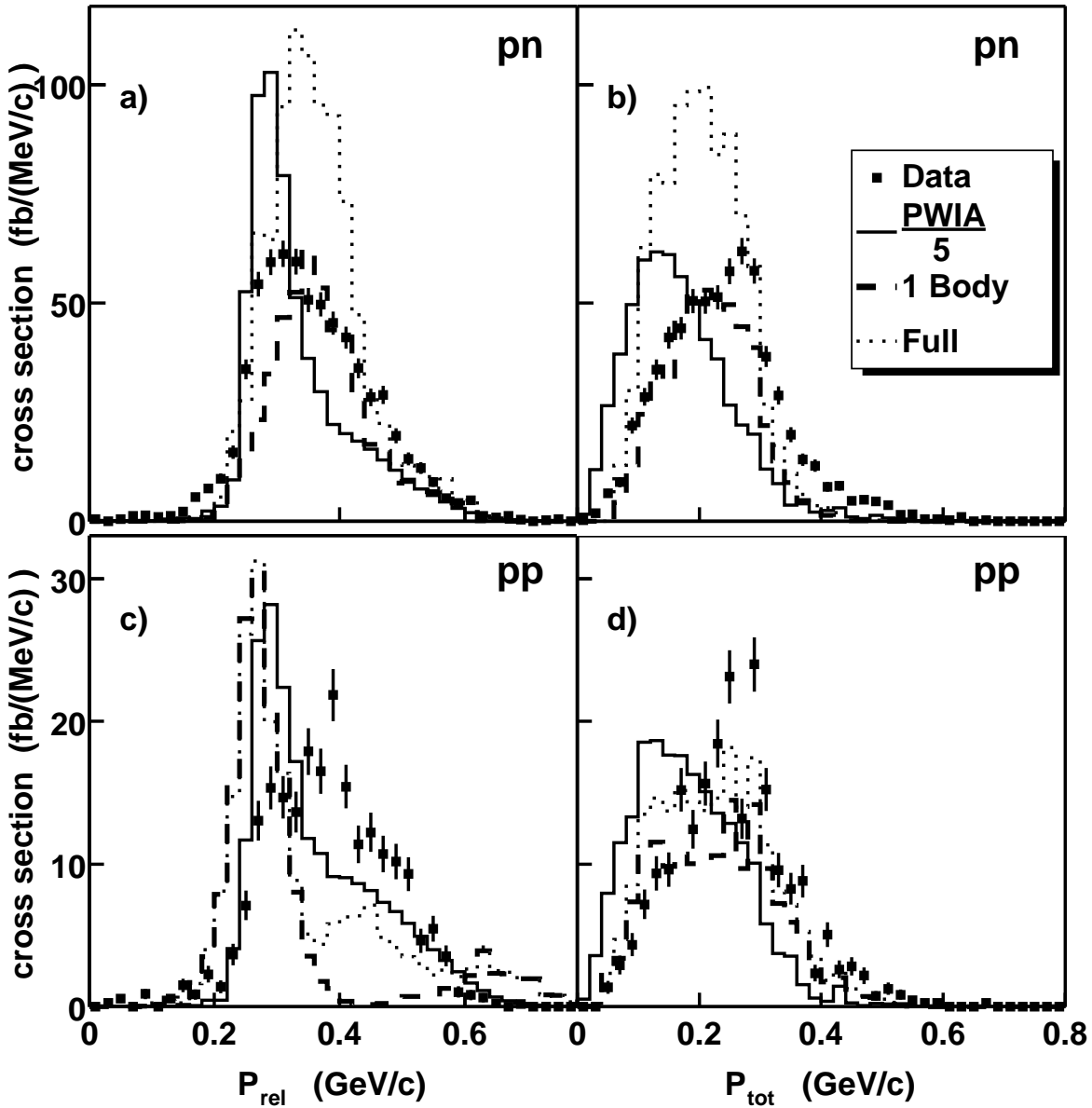}
\caption{\label{src:spectator}
Figure a) shows lab frame cross section versus pn pair relative momentum.
Points show the data, solid histogram shows the
PWIA calculation times $\frac{1}{5}$, dashed histogram shows
Laget's one-body calculation~\cite{Laget:1987jr,Audit:1996tq,Laget:1987zz}, dotted histogram shows
Laget's full calculation; b) the same for total momentum;
c), d) the same for pp pairs.
Reprinted with permission from 
\href{http://dx.doi.org/10.1103/PhysRevLett.92.052303}{Niyazov R~A {\em et~al.\/} (CLAS) 2004 {\em Phys. Rev. Lett.\/} {\bf 92} 052303.}
Copyright 2004 by the American Physical Society.}
\end{figure}

Instead of directly knocking out one nucleon of the correlated pair, 
it is also possible in the three-body system to knock out 
the third nucleon and observe the decay of the spectator correlated pair.  
This was done in the $^3$He$(e,e'pp)n$ using 2.2 and 4.4 GeV electron beams 
and detecting the scattered electron and ejected protons in CLAS over a 
wide kinematic range~\cite{Niyazov:2003zr}.  When all three final state nucleons have 
momenta greater than 250 MeV/c, the reaction is dominated by events 
where two nucleons each have less than 20\% of the transferred energy 
and the third 'leading' nucleon has the remainder.  Final state interactions 
of the leading nucleon are suppressed by requiring that it has perpendicular momentum 
with respect to $\vec q$ of less than 300 MeV/c.  In these cases the two other 
nucleons (the $pn$ or $pp$ pair) are predominantly back-to-back and have very 
little total momentum in the momentum transfer direction.  The relative pair momentum 
is also mostly isotropic.  This indicates that the nucleon-nucleon pair is  correlated and 
is a spectator to the virtual photon absorption.  This means that the measured 
relative and total pair momenta, $\vec p_{rel} = (\vec p_1 - \vec p_2)/2$ 
and $\vec p_{tot} = \vec p_1 + \vec p_2$  are closely related to the initial momenta 
of the correlated pair.  As shown in~\fref{src:spectator}, the pair relative momentum peaks 
at about 300--400 MeV/c and the pair total momentum peaks at about 300 MeV/c.  
The $pp$ and $pn$ momentum distributions are very similar.  The advantage of 
this approach is that there is no contribution from meson exchange currents 
or isobar configurations since the virtual photon does not interact with 
the correlated pair.  However, because the continuum interaction of the correlated 
pair is very strong and reduces the calculated cross section by a factor of about 
ten, this reaction is very difficult to calculate precisely.  Diagrammatic 
calculations by Laget are in qualitative agreement with the measured momentum 
distributions, but only after including the effects of the continuum interaction.

\section{Summary and Outlook}

From these new results, one can get a good general
picture of two-nucleon correlations that take place within the nucleus.
From $(e,e^{\prime})$ we learn the probability of correlations in various nuclear systems.
From $(e,e^{\prime}p)$ we learn the probability of independent single particle liked knock-out 
and about the high-momentum tail in the measured momentum distributions.
From $^3$He(e,e'pp)n we learn the about shape of the distorted correlated 
pair momentum distributions and from $^{12}$C$(e,e^{\prime}pN)$ we 
learn the relative importance of $pp$, $pn$ and $nn$.
Putting this all together, we learn that in $^{12}$C 60-70$\%$ of the nucleons within the nucleus
act like independent particle, while approximately 20\% have a short-range correlation with another
nucleus and that these partner nucleon almost always have the opposite isospin due to
short-range tensor correlations.  The remaining fraction of nucleons within the nucleus 
are thought to be in long-range correlations.

The association of the small
$(e,e^{\prime}pp)$/$(e,e^{\prime}pn)$
ratio with the dominance of the nucleon-nucleon short-range tensor
force leads naturally to the quest to increase the missing momentum
and to look for pairs which are even closer to each other, at
distances that are dominated by the repulsive core.
In particular, the observed dominance of $np$ to $pp$ correlations is predicted to decrease as
missing momentum is increased and the nucleon-nucleon short-range 
tensor force gives way to the short range repulsive force~\cite{Schiavilla:2006xx,Sargsian:2005ru,Alvioli:2007zz,Wiringa:2008dn}.
This momentum dependence
is also expected to be more pronounced in light nuclei, motivating an
upcoming Hall~A triple coincidence experiment, similar to the $^{12}$C$(e,e^{\prime}pN)$ measurement,
which will map out the $np$ to $pp$ ratio over the range $400 < p_{miss} <
850~{\rm MeV}/c$ using ${}^4{\rm He}$.

Further $(e,e^{\prime})$ $x_B > 1$ inclusive measurements are also
planned.  In the near term, Hall~A will make a comparison of symmetric and
neutron rich isotopes of Calcium to probe the isospin dependence of
$2N$ and $3N$ correlations.  This experiment will also make a
detailed study of the onset of scaling due
to $3N$ correlations will be made from the $Q^2$ dependence of $x_B > 2$
cross sections.   After the 12~GeV upgrade, Hall~C will measure the $(e,e^{\prime})$ reaction
in $x_B > 1$ kinematics  and
in the deep inelastic scattering region of $W > 2$~GeV.
This will allow an investigation of quarks which are being shared by more 
than one nucleon~\cite{Sargsian:2002wc,Fromel:2007rc}.

Beyond the approved experiments, future facility upgrades could include a large
acceptance detector to detect backward recoil particles in coincidence with
forward angle spectrometers.
Such a detector system would allow broader studies of $(e,e^{\prime}pN)$ reactions and
the detection of the products of $3N$ or $N\Delta$ correlations.

\ack

Many thanks to theorists Mark Strikman and Misak Sargian for the their invaluable help in teaching us the physics
of correlations and  in preparing this manuscript;
and well as thanks to theorists Claudio Ciofi degli Atti, Leonid Frankfurt, Jean-Marc Laget, Rocco Schiavilla, 
and many others for thought provoking discussions and help in understanding the results presented herein. 
Thanks to experimentalists Stepan Stepanya, Shalev Gilad, Larry Weinstein, William "Bill" Bertozzi, John Watson, 
John Arrington, Donal Day, and many others who have helped push forward the short-range correlations program 
at Jefferson Lab.  But most of all, our thanks goes to the people who received 
their Ph.D.'s doing the research presented herein and
without whom these exciting new results would not have been possible.
In Hall~A, thank to Nilanga Liyanage for his work on the E89-003 $^{16}$O$(e,e^{\prime}p)$ experiment;
to Fatiha Benmokhtar and Marat Rvachev for their work on the E89-044 
$^{3}$He$(e,e^{\prime}p)$ 
experiment; and to Peter Monaghan, Ran Shneor, and Ramesh Subedi for their work on the E01-015 $^{12}C(e,e'pN)$ experiment.
In Hall~B, thanks to Natalia Dashyan for her work on the CLAS (e,e') analysis 
and Rustam Niyazov for his work on the CLAS $^3$He(e,e'pp)n analysis.
In Hall C, thanks to Ben Clasie, Dipangkar Dutta, Dave McKee, Kristoff Normand, Daniela Rohe, 
and Derek van Westrum for their work on nuclear transparency and spectral function measurements.
And finally, we remember our deceased friend Kim Eyigan, who shorty before his death, 
found the signatures for short-range correlations that he had spent his life searching for.


\section*{References}


\providecommand{\newblock}{}

\end{document}